# Li DOPING EFFECT ON PROPERTIES AND PHASE TRANSFOMATIONS of $KNbO_3$.

V. A. Trepakov[1,2,*], M. L. Savinov[1], V Železný[1], P. P. Syrnikov[2],

A. G. Deyneka[1] and L. Jastrabik[1]

[1]Institute of Physics ASCR, 182 21 Prague 8, Czech Republic

[2]A.F. Ioffe Physico-Technical Institute, 194021, St.-Petersburg, Russia

**ABSTRACT**. Dielectric permittivity and infrared reflectivity spectra of Li doped $KNbO_3$ single crystals have been studied for the first time for $K_{1-x}Li_xNbO_3$ (KLN) with $x = 0.015$, 0.02, 0.065. It was found that like in $KTaO_3$, Li admixture results in appearance of dielectric relaxation with the relaxation parameters very close to those in $KTaO_3$ quantum paraelectric. It was attributed to $\pi/2$ dipole reorientation of $Li^+$ <100> off centers substituted $K^+$, which appear to be presents as in paraelectric cubic phase as in ferroelectric phase down to low temperatures. Besides, Li doping is accompanied by increasing of the cubic-tetragonal phase transition point, decreasing of tetragonal-orthorhombic-rhombohedral phase transition points and TO soft mode stiffen at room temperature.

INTRODUCTION AND BACKGROUND.

Doping with Li, which substitutes K in the $KTaO_3$ (KTO) quantum paraelectric and forms $Li^+$ off-center ions with large (~1.2 Å) displacement along <100> directions has been subject of extensive studies for a long period of time (see, e.g. [1-5] and references therein). $Li^+$ ions substitute for $K^+$ taking six off-center positions in the potential wells along the <100> directions. Such doping crucially modifies $KTaO_3$ properties, leading to appearance of a large dielectric relaxation polarization and various low-temperature multi-scale dipole ordering effects, form polar glass to long range ordered polar state, depending on Li conce-

E-address: trevl@fzu.cz; trevli@MS8484.spb.edu

ncentration. At the same time, Li doping effect in related ferroelectric $KNbO_3$ (KNO) is not known. Presumably, only in [6] some physical-chemical aspects of $KNbO_3 - LiNbO_3$ system using ceramic species had been reported. It was found that at low Li concentration $K_{1-x}Li_xNbO_3$ (KLN) has perovskite-type structure with stability limit $x_s \sim 10\text{-}15\%$ mol. and Li admixture decreases dielectric permittivity and piezoelectric response. At the same time there are serious fundamental science and technological reasons in studies of KLN. Firstly, potassium niobate as $BaTiO_3$ is the well known model ferroelectric, obeying three cubic-tetragonal- orthorhombic-rhombohedral phase transitions (PT) of the first order at 703 $K^h$, 495 $K^h$ and 221 $K^c$ respectively ($c-$ denotes cooling, $h$- heating run), keeping orthorhombic symmetry from -50$^0$C to 222$^0$C [7]. It has the largest effective nonlinear optical coefficient among the commonly used inorganic crystals. Therefore, it is normally used for frequency doubling low and medium power lasers such as diode lasers, diode pumped solid state lasers and cw Nd:YAG and Ti:Sapphire lasers. It is also an efficient crystal for low oscillation threshold and high efficient optical parametric oscillators (OPO) and generators (OPG). However, heating over 210$^0$C and cooling below -40$^0$C can destroy the crystal as well as indelicate no smoothly operation with it and temperature instability lead to domain structure appearance and instable output, which limits KNO applicability for practice. We believe that Li doping can sufficiently modify these properties, and improve KNO crystals stability. Secondly, recent finding demonstrated that related lead-free (K,Na)$NbO_3$-$LiTaO_3$ materials evidence very promising piezoelectric properties ($d_{33}$ is up to $\sim 230$ pCN$^{-1}$) [8]. At last, KLN is the limit case for of $K_{1-x}Li_xTa_{1-y}Nb_yO_3$ solid solutions, are known by intriguing properties and phase transitions as well as prospective applications in electrically controlled holographic and compositionally graded pyroelectric devices [9,10]. We report on the first synthesis and studies of the KLN crystals extending investigations of Li doping effect on properties and phase transformations of the classical ferroelectric assuming that in KLN, small Li$^+$ ions (ion radius of ~ 0.68 Å) also substitute K$^+$ (1.64 Å [11] lattice sites and form strong reorientable <100> off dipole centers. It was intended: *i)* to perform growth KLN single crystals; *ii)* to inspect ability of Li admixture, Li$^+$ related dipole <100> centers to modify phase transitions of KNO, and maybe even suppress some transitions; *iii)* to show up presence (*or absence*) of Li$^+$ related dielectric relaxation in paraelectric cubic as well as in ferroelectric phases and at low temperatures. The main attention was paid to Li doping effect on cubic-tetragonal-rhombic phase transitions.

**EXPERIMENTAL**.

$K_{1-x}Li_xNbO_3$ single crystals with x = 0.015, 0.02 and 0.065 (KLN-1.5, KLN-2, and KLN-6.5) were grown by spontaneous polarization from slowly cooled flux The batch consisted of $Nb_2O_5$, with approximately a 15% excess of $K_2CO_3$ and certain amount of $Li_2CO_3$ ultra high purity. Obtained crystals were transparent with resistivity exceeded $10^{10}$ Ω cm. The Li content inside the experimental specimens was determined by AAS analysis using a VARIAN AAS Spectrometer AA-30. X-ray analysis confirms perovskite-like structure for all KLN-1.5, KLN-2 and KLN-6.5 crystals. Experimental specimens were fabricated as thin platelet condensers with surfaces oriented normal to the [100] principal axis of the paraelectric cubic phase. The complex dielectric permittivity was studied in the 100 Hz – 1 MHz frequency range and 10- 800 K temperature interval using HP 4192A impedance analyzer. Complementary non-polarized far-infrared reflectivity was measured at RT with a Bruker IFS 113v spectrometer in the range 20-2000 cm$^{-1}$ and fitted using the factorized form of dielectric function [12].

**RESULTS UAND DISCUSSION:**

All compositions under study reveal very weak dielectric permittivity dispersion is caused by presence of domains in ferroelectric phases and conductivity increasing in paraelectric cubic phase. In each the case presence of the three characteristic ε′(T) maximums and temperature hysteresis evidenced three first order transitions inherent to parent KNbO₃. Figure 1 presents the dielectric constant for KLN-1.5 is taken on cooling run for the temperature region of cubic-tetragonal- rhombohedral PT and respective PT points for KNO [7].

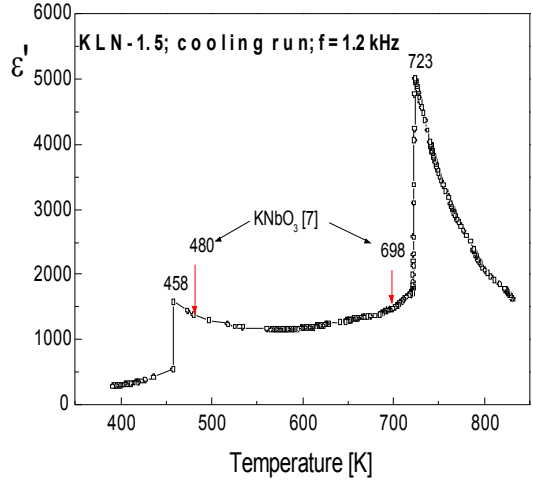 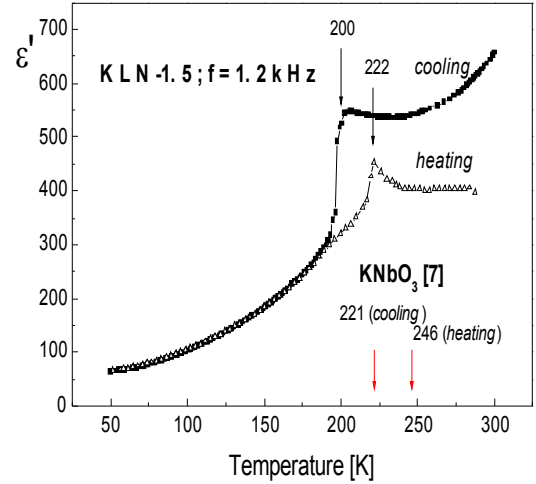

Figure 1. Dielectric permittivity for KLT-1.5 in the region cubic-tetragonal-orthorhombic phase transitions; arrows show PT points for KNbO₃.

Figure 2. Dielectric permittivity for KLT-1.5 in the region of orthorhombic-rhombohedral phase transition.

It is seen that Li admixture leads to increasing of the cubic-tetragonal transition point and decreasing of the tetragonal-orthorhombic one, i.e. broadens temperature region of the tetragonal phase stability. Magnitudes of the $\varepsilon'(T)$ maximums appeared to be larger that those in KNO in the vicinity of cubic-tetragonal PT (~ 4500) and nearly the same for tetragonal-orthorhombic one [13]. Figure 2 presents dielectric permittivity for KLN-1.5 in the orthorhombic-rhombohedral transition region. It is seen that Li admixture decreases temperature point of this transition. The same PT shifts were found in KLN-2 and KLT-6.5 with magnitude of shifts increasing with Li concentration. So, position of $\varepsilon'(T)$ maxima appeared to be in KTN-6.5 ~740 $K^c$ and 747 $K^h$ for cubic-tetragonal and 425 $K^c$, 453 $K^h$ for tetragonal-orthorhombic phase transitions. Figure 3 represents the dielectric relaxation was found in KLT-1.5 at low temperatures manifesting in the tanδ(T) maximum shift to higher temperatures with frequency. We found that the relaxation satisfactorily obeys an Arrhenius law $\tau = \tau_0 \exp(U/kT)$ with $\tau_0 \approx 2 \times 10^{-13}$ s and $U \approx 94.7$ meV (~1100 K), which is very close to the $Li^+$ relaxation parameters in $KTaO_3$. Figure 4 presents the non-polarized IR reflectance spectrum of KLT-2 is taken at RT and fit by factorized oscillator model. Reflectance spectrum of the parent cubic paraelectric phase contains three characteristic restrahen bands for perovskites in the region ~ 60- 600 $cm^{-1}$ which are split by symmetry change due to phase transitions [14]. As result at room temperature KLN-2 is in the orthorhombic phase and has three IR active irreducible representations $A_1$, $B_1$, $B_2$. In comparison with pure KNO the reflectivity of KLN-2 in the range of soft mode is much lower (KNO ~85%; KLN - ~62%). It leads to two important suggestions: *i*) Li admixture stiffen TO soft mode in KNO at RT, *ii*) lattice modes provide contribution into low-frequency permittivity yield only $\varepsilon' \sim 70$ (from IR) in contrast with low frequency $\varepsilon' \sim 290$ (dielectric permittivity), which clearly indicates presence of relaxation contribution connect with $Li^+$ related dielectric relaxation.

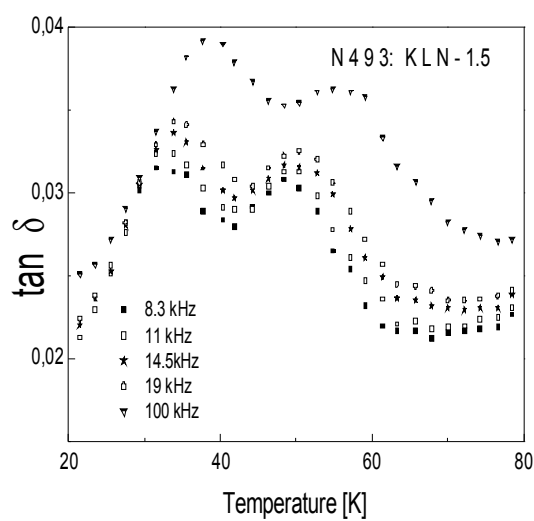 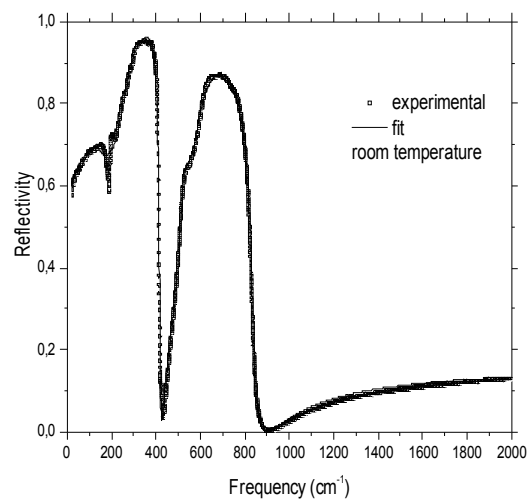

Figure 3. Low temperature dielectric losses.   Figure 4. IR reflectivity and fit by factorized oscillator model for KLT-2

**CONCLUSION.**

It was shown that Li doping of KNO crystals leads to dielectric permittivity decreasing at RT due to TO soft mode stiffness. Magnitude of the low frequency permittivity temperature maximum at the cubic-tetragonal phase transition increases that is connected with $Li^+$ related dielectric relaxation contribution. Like in KTO, $Li^+$ presents in KNO as <100> reorientable dipole off-centers, and not only in paraelectric, but also in deep ferroelectric phases at low temperatures. Relaxation parameters obtained at low temperatures in the orthorhombic phase appeared to be very close to those in KTO, that means that FE ordering of KNO not strongly influences $Li^+$ relaxation characteristics.

Moderate concentrated KLN, similarly as the parent $KNbO_3$, evidences three characteristic first order phase transitions. Li admixture appreciably influences position of the phase transitions: temperature of the cubic-tetragonal ferroelectric PT increases and temperature of the tetragonal-rhombic PT decreases (tetragonal phase stabilization), temperature of the orthorhombic-rhombohedral PT decreases as well. These shifts increase with Li content.

**ACKNOWLEDGEMENTS.**

This was supported by Grants 1QS100100563 AS CR**,** AV ČR AV0Z 10100522 and RFBR 06-02-17320.

REFERENCES:

1.  Höchli, U. T., Knorr, K. & Loidl, A., Orientation glasses. Adv. Phys.1990, **39,** 405-615.

2.  Prosandeev, S. A., Trepakov, V. A., Savinov, M. E. & Kapphan, S. E. Coupling of $Li^+$ relaxators to the soft mode in $KTaO_3$:Li.", J. Phys.: Condens. Matter 2001, **13,** 719-725.

3.  Prosandeev, S. A., Trepakov, V. A., Savinov, M. E., Jastrabik, L., & Kapphan, S. E. Characteristics and the nature of the low-frequency dielectric response in moderately concentrated $KTaO_3$:Li.", J. Phys.: Condens. Matter, 2001, **13,** 9749-9760.

4. Prosandeev, S. A. & Trepakov, V. A. Dielectric response in quantum paraelectrics containing dipole impurities., JETP 2002, **94,** 419-430.

5. Tupitsyn, I. I., Dejneka, A. G., Trepakov, V. A., Jastrabik, L. & Kapphan, S. E. Energy structure of $KTaO_3$ and $KTaO_3$:Li., Phys. Rev. B 2001, **64,** 195111-1 – 195111-6.

6. Dambekalne, M. Ya., Yanson, G. D.& Freidenfeld, E. Zh..Production processes and physico-chemical properties of solid solutions in $KNbO_3$-$LiNbO_3$ system.", Procs. of the 3rd Inter-industrial conference on fabrication methods and analysis of the ferroelectric and piezoelectric materials and sources., Edit. of the All-Union Inst. of Reactive and Chemical Pure Materials for Electronic Technique. Part I, Donezk, 1970, 68 – 78.

7. Cotts, R. M. & Knight, W. D., Nuclear Resonance of $Nb^{93}$ in $KNbO_3$.Phys. Rev. 1954, **96**, 1285–1293.

8. Saito, Y., Takao, H., Tani, T., Nonoyama, T., Takatori, K., Homma, T., Nagaya, T., & Nakamura, M. Lead-free piezoceramics., Nature 2004, **432**, 84-87.

9. Trepakov, V., Savinov, M, Giulotto, E., Galinetto, P., Camagni, P., Samoggia, G., Boatner, L. A. & Kapphan, S. E. Dipole ordering effects and reentrant dipolar glass state in $KTaO_3$:Li,Nb., Phys. Rev. B, 2001, **63** 172203-1 – 172203-4.

10. Trepakov, V. A., Jastrabik, L., Kapphan, S., Giulotto, E. & Agranat, R., "Phase transitions, related properties and possible applications of $(K,Li)(Ta,Nb)O_3$ crystals.", Optical Materials, 2002, **19**, 13-21.

11. Shannon, R. D. & Prewitt, C. T., Effective ionic radii in oxides and fluorides. Acta. Cryst. 1969, **B25**, 925-946.

12. Gervais, P., *High-temperature Infrared Reflectivity Spectroscopy by Scanned Interferometry.* New York: Academic Press; 1983. Chapter in a Book: K. J. Button, Infrared and Millimeter Waves. Vol. **8**, p. 280.

13. Shirane, G., Danner, H., Pavlovic, A., & Pepinsky, R., Phase transitions in ferroelectric KNbO$_3$., Phys. Rev. 1954, 93, 672-673.

14. Servoin, J. L., Gervais, F., Quittet, A. M. & Luspin Y., Infrared and Raman responses in ferroelectric perovskite crystals: Apparent inconsistencies. Phys. Rev. 1980, B 21, 2038–2041.